\documentclass[twocolumn]{aastex701}

\newcommand{\mstar}{M$_\ast$}
\newcommand{\muv}{$M_{\rm UV}$}

\begin{document}

\title{Little Red Dots as Globular Clusters in Formation}

\author[0000-0002-0302-2577]{John Chisholm}
\email{chishom@austin.utexas.edu}
\affiliation{Department of Astronomy, The University of Texas at Austin, Austin, TX 78712, USA}
\affiliation{Cosmic Frontier Center, The University of Texas at Austin, Austin, TX 78712, USA} 
\correspondingauthor{John Chisholm}


\author[0000-0002-4153-053X]{Danielle A. Berg}
\email{daberg@austin.utexas.edu}
\affiliation{Department of Astronomy, The University of Texas at Austin, Austin, TX 78712, USA}
\affiliation{Cosmic Frontier Center, The University of Texas at Austin, Austin, TX 78712, USA} 

\author[0000-0002-9604-343X]{Michael Boylan-Kolchin}
\email{mbk@astro.as.utexas.edu}
\affiliation{Department of Astronomy, The University of Texas at Austin, Austin, TX 78712, USA}
\affiliation{Cosmic Frontier Center, The University of Texas at Austin, Austin, TX 78712, USA} 
\affiliation{Weinberg Institute for Theoretical Physics, The University of Texas at Austin, Austin, TX 78712, USA}

\author[orcid=0000-0002-2380-9801]{Anna de Graaff}\thanks{Clay Fellow}
\email{degraaff@mpia.de}
\affiliation{Center for Astrophysics, Harvard \& Smithsonian, 60 Garden St, Cambridge, MA 02138, USA}
\affiliation{Max-Planck-Institut f\"ur Astronomie, K\"onigstuhl 17, D-69117 Heidelberg, Germany}

\author[orcid=0000-0001-6278-032X]{Lukas J. Furtak}
\email{furtak@utexas.edu}
\affiliation{Department of Astronomy, The University of Texas at Austin, Austin, TX 78712, USA}
\affiliation{Cosmic Frontier Center, The University of Texas at Austin, Austin, TX 78712, USA}

\author[0000-0002-5588-9156]{Vasily Kokorev}
\email{vkokorev@utexas.edu}
\affiliation{Department of Astronomy, The University of Texas at Austin, Austin, TX 78712, USA}
\affiliation{Cosmic Frontier Center, The University of Texas at Austin, Austin, TX 78712, USA} 
 
\author[0000-0003-2871-127X]{Jorryt Matthee}
\email{jorryt.matthee@ist.ac.at}
\affiliation{Institute of Science and Technology Austria (ISTA), Am Campus 1, 3400 Klosterneuburg, Austria}

\author[0000-0002-8984-0465]{Julian B.~Mu\~noz}
\email{julianbmunoz@utexas.edu}
\affiliation{Department of Astronomy, The University of Texas at Austin, Austin, TX 78712, USA}
\affiliation{Cosmic Frontier Center, The University of Texas at Austin, Austin, TX 78712, USA} 
\affiliation{Weinberg Institute for Theoretical Physics, The University of Texas at Austin, Austin, TX 78712, USA}

\author[0000-0003-3997-5705]{Rohan P.~Naidu}
\email{rnaidu@mit.edu}
\altaffiliation{NASA Hubble Fellow, Pappalardo Fellow}
\affiliation{MIT Kavli Institute for Astrophysics and Space Research, 70 Vassar Street, Cambridge, MA 02139, USA}

\author[0000-0002-2090-9751]{Andreas A.C. Sander}
\email{andreas.sander@uni-heidelberg.de}
\affiliation{Zentrum für Astronomie der Universität Heidelberg, Astronomisches Rechen-Institut, Mönchhofstr. 12-14, 69120 Heidelberg}
\affiliation{Interdisziplinäres Zentrum für Wissenschaftliches Rechnen, Universität Heidelberg, Im Neuenheimer Feld 225, 69120 Heidelberg, Germany}

\begin{abstract}
Little Red Dots (LRDs), among the most enigmatic high-redshift discoveries by JWST, are commonly believed to be powered by accreting supermassive black holes. Here, we explore the possibility that these sources are globular clusters in formation, with rest-frame UV arising from a very young stellar population and rest-frame optical from a short-lived supermassive ($>10^4$~M$_\odot$) star. The spectral profiles of LRDs are broadly consistent with this scenario, though the observed temperatures and bolometric luminosities favor emission reprocessed by optically thick, continuum-driven winds not fully captured by current models. The LRD $z\sim5-7$ UV luminosity function naturally evolves, under standard evolutionary and mass-loss prescriptions, into a present-day mass function with a turnover at $\log_{10}($\mstar/$M_\odot)=5.3$ and an exponential cutoff at high masses, consistent with local globular-cluster populations. We estimate the total present-day number density of LRDs formed across all redshifts to be $\approx0.3$~Mpc$^{-3}$, similar to local globular clusters. The observed LRD redshift range matches the age distribution of metal-poor globular clusters, without current LRD counterparts to the metal-rich population. If LRDs are globular clusters in formation, we predict chemical abundance patterns characteristic of multiple stellar populations, including enhanced He and N, and potential Na-O and Al-Mg anti-correlations. These results offer a local perspective to explore this surprisingly abundant population of distant sources, and a potential new window into extreme stellar astrophysics in the early Universe.
\end{abstract}


\section{Introduction} 
Little Red Dots \citep[LRDs; ][]{Matthee} are one of the most enigmatic populations revealed by JWST at high redshift. The discovery of a cosmically abundant population of compact red sources at $z > 3$, largely missed by previous telescopes, quickly captivated the community \citep{labbe23}. The LRD spectral shape became their hallmark: a pronounced V-shape, with a blue rest-frame UV continuum and red rest-frame UV-to-optical. The proximity of this inflection point to the Balmer break \citep{wang24} led some to conclude that LRDs were a population of massive evolved galaxies, which, if confirmed, would have pushed $\Lambda$CDM galaxy formation to uncomfortable places \citep{mbk23}. While compelling, the star-forming galaxy scenario faced significant problems \citep{endsley23, sabti}, most notably lensed LRDs have extremely small sizes $<30$~pc \citep{furtak23, Yanagisawa}  and many LRDs lack strong infrared dust emission despite their red rest-frame optical colors \citep{perez-gonzalez, akins25a, leung, setton25a}. 

The discovery of broad ($\gtrsim1000$~km~s$^{-1}$) H$\alpha$ emission provided a concrete clue for the origin of LRDs \citep{harikane23, kocevski, Matthee}. Interpretations invoking accreting black holes quickly came to dominate their discussion \citep{kocevski, maiolino, greene25}. If LRDs are powered by accreting black holes, they are remarkably distinct from other known black hole populations. LRDs rarely have detected X-rays \citep{Ananna, juodzbalis}, many exhibit Balmer absorption even in H$\alpha$ \citep{Matthee}, lack the rest-frame mid-IR colors typical of a dusty torus \citep{perez-gonzalez}, occupy lower mass halos than standard quasars \citep{pizzati}, and have inferred black-hole-to-stellar-mass ratios approaching unity \citep{furtak, maiolino}. Individually, these properties can each be explained as black holes forming before the bulk of stars by accreting at super-Eddington rates \citep{lambrides} with their accretion disks enshrouded by dense neutral gas cocoons \citep{Inayoshi, naidu25}. Since accreting black holes are expected to evolve strongly with redshift, the extreme accretion physics powering the LRD population was viewed as a natural extension of early black hole formation. 

The discovery of \lq{}\lq{}Black Hole Stars" \citep{naidu25, degraff_cliff}, which resemble LRDs without significant UV continua, further highlights their unusual properties. These objects suggest the characteristic LRD V-shape can be \textit{spectroscopically} decomposed into two distinct components: one in the rest-frame UV and one in the rest-frame optical \citep[][]{barro26, wendy}. The rest-frame UV is well-described by a young stellar population, while the rest-frame optical is well-fit by a modified blackbody with temperatures 2000-7000~K and luminosities 10$^{9-11}$~L$_\odot$ \citep{degraaf}. The spectral inflection occurs due to the relative strength of the UV and optical components, shifting as red as $\sim6000$~\AA\ \citep{degraaf}.

Standard sub- or super-Eddington accretion flow models produce multi-temperature spectra with significant EUV/X-ray emission \citep{shakura73, pringle81, abramowicz88, sadowski, lambrides}.  If accretion powers LRDs, this intrinsic high-energy radiation must be reprocessed by an optically thick medium. The emergent spectrum must be converted into a single-temperature \textit{cool} modified blackbody while suppressing strong multi-wavelength radiative signatures of cooling and heating: X-ray emission, radio free-free emission, prominent metal cooling lines, and dust. The observed Planck spectrum further implies that the broad emission lines cannot originate deep within fully thermalized layers; instead they must arise in gas exterior to the photosphere, possibly in a stratified atmosphere. At these cool temperatures the photosphere is largely neutral, implying that Balmer emission is likely collisionally excited in the atmosphere \citep{torralba}. The observed single-temperature blackbodies place stringent radiative-transfer and energetic constraints of LRDs. 

Here, we explore an alternative scenario in which LRDs are globular clusters in formation \citep[e.g.,][]{gieles18}. In this scenario, the LRD V-shape arises from an ultra-young stellar population producing the rest-frame UV continuum, while a supermassive star (SMS) produces the rest-frame optical. Young stellar clusters are extremely dense \citep{adamo, Claeyssens, Claeyssens26} with high stellar multiplicity that suffer frequent dynamical interactions \citep{offner} that can drive runaway stellar mergers to form SMSs, even in relatively metal-rich environments \citep{gieles18, vergara}. Thus, SMS formation may naturally occur in dense stellar environments to resolve both the globular cluster mass-budget problem \citep{Renzini} and the origin of multiple populations \citep{carretta, Piotto, gratton10, bastian, milone}. 

SMSs remain largely theoretical \citep{Kuruvanthodi}, but they may reproduce many LRD features \citep{martins, furtak23, nandal}. They are theorized to have masses of 10$^{4-6}$~M$_\odot$ with luminosities near $10^{7-11}$~L$_\odot$ \citep{gieles18, martins, nandal}. Their extreme luminosities drive dense winds that can mimic the broad H$\alpha$ observed in LRDs \citep{martins} through continuum opacities that include electron scattering and H$^-$ \citep{dotan, Denissenkov, vink18}. SMSs are expected to be large ($1,000$~au) leading to relatively cool temperatures, $T_{\rm eff} \sim 3000-10,000$~K \citep{haemmerle, haemmerle19, martins, Bernini-Peron}. The immense gravitational potentials of SMSs require a continuous accretion of fresh hydrogen to stave off collapse to a black hole \citep{begelman, woods}, and SMSs likely only exist for order of magnitude $\sim$1~Myr \citep{portegies02, gieles18}. SMSs are predicted to have a \lq{}\lq{}conveyor-belt"-like fully-convective structure \citep{gieles18} that may explain the unique odd-element enhancements observed in local globular clusters including Na-O and Al-Mg anti-correlations, as well as He and N enhancements \citep{carretta, Piotto, Gratton, Denissenkov, bastian, milone, gieles25}. 

Here, we critically examine this hypothesis by envisioning that LRDs are a short-lived phase of globular cluster formation during which both an SMS and a stellar cluster coexist. We first demonstrate that the observed V-shapes profiles are plausible with a globular cluster hosting an active SMS (\autoref{spec}) and explore the limitations of current SMS models lacking sufficiently optically thick winds (\autoref{sms}). We then use the observed LRD UV luminosity function to estimate the $z\sim7$ LRD mass function and passively evolve it to $z=0$ (\autoref{LRDs}). We compare the shape (\autoref{shape}) and number counts (\autoref{counts}) of this mass function to local globular clusters (\autoref{GCs}). Finally, in \autoref{discuss} we present three testable predictions for the hypothesis that LRDs are globular clusters in formation. 

Throughout this paper we use a standard $\Lambda$CDM cosmology with $H_0 = 67.4$, $\Omega_\Lambda = 0.685$, and $\Omega_{\rm m} = 0.315$ \citep{planck}. All magnitudes are AB-magnitudes. All logarithms are base 10.

\section{Globular Clusters in Formation with a Supermassive Star: V-shaped Spectral Profiles}
 \label{spec}
\begin{figure}
    \centering
    \includegraphics[width=\linewidth]{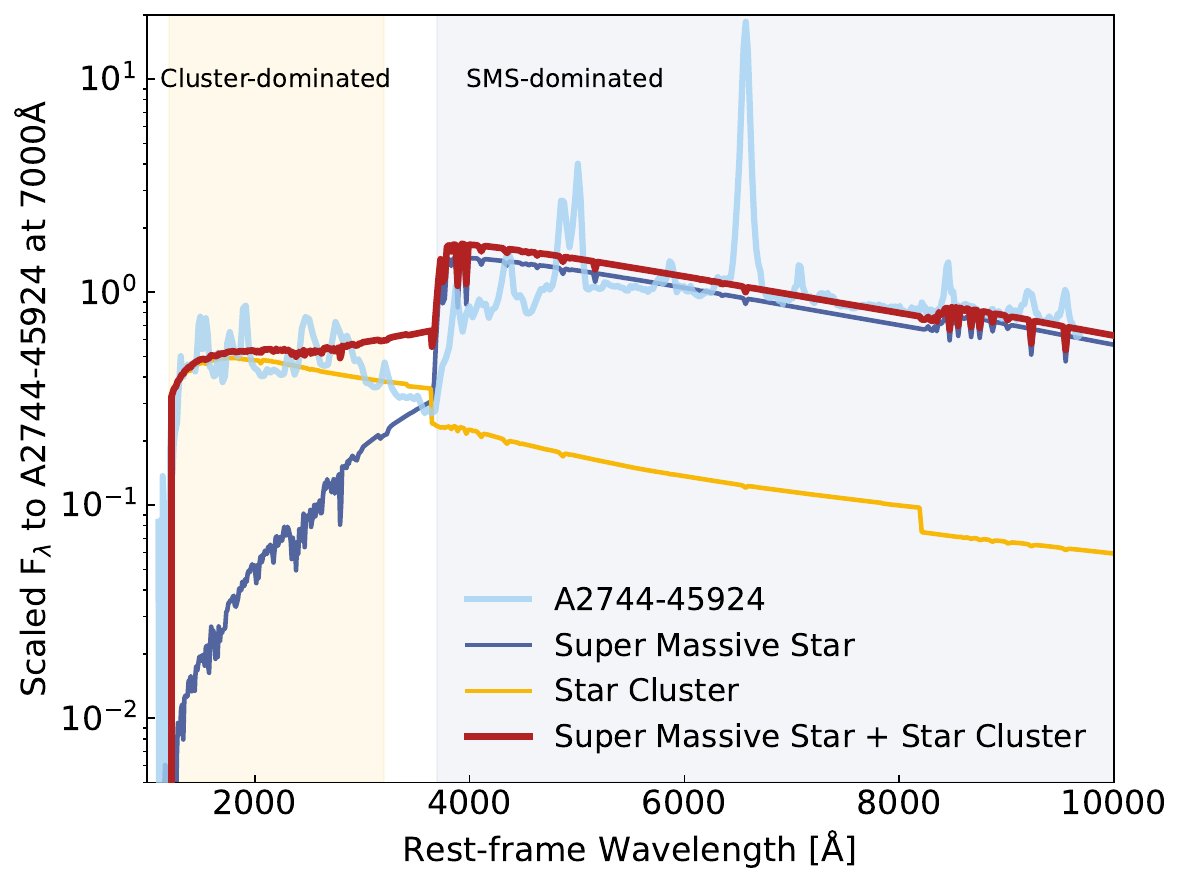}
    \caption{Illustration of a bright LRD A2744-45924 \citep[light-blue line;][]{labbe24} as a globular cluster in formation \citep{martins}. The gold line shows a young star cluster with nebular continuum self-consistently added, while the dark-blue line shows a supermassive star model. The combined shape (red line) highlights a \lq{}\lq{}V-shaped\rq{}\rq{} morphology as the star cluster dominates the UV and the supermassive star dominates the optical. The supermassive star model is not tuned to match the spectrum. Cooler supermassive star models may reduce the mismatch near 3500~\AA, but no such models currently exist (see \autoref{sms}).}
    \label{fig:spec}
\end{figure}

LRDs are commonly defined by their V-shaped spectral morphology. In \autoref{fig:spec}, we show A2744-45924 \citep[light-blue spectrum; ][]{labbe24}, one of the hottest and most luminous known LRDs \citep[$T_{\text{eff}} \sim 5700$~K; ][]{torralba}. This source is particularly well-suited to compare to publicly available SMS models with $T_{\rm eff} = 7,000-10,000$~K \citep{martins}. We adopt the coolest SMS model from \citet{martins}, \lq{}\lq{{}A2\rq{}\rq{}, with $T_{\rm eff} = 7000$~K, $L_{\rm SMS} = 10^9$~L$_\odot$, and $M_{\star}=53{,}956$~M$_\odot$. The cluster component is a young stellar population with a Kroupa initial mass function \citep{kroupa} that has nebular continuum self-consistently added \citep{martins}. To match A2744-45924, we include damped Ly$\alpha$ from the foreground IGM at $z=4.46$ \citep{inoue14}, and fit a single, uniform dust screen using the SMC attenuation law \citep{Pei}. The red curve in \autoref{fig:spec} shows the best-fit combined cluster plus SMS model, with $E(B-V) = 0.25$~mag ($A_{1500} = 3$~mag and $A_V = 0.69$~mag).

An SMS combined with a star cluster can plausibly reproduce the observed V-shaped spectral profile. The UV is dominated by a very young stellar population, while the SMS naturally produces the appearance of a strong Balmer break and red rest-frame UV-to-optical continuum \citep{martins}. The transition between an SMS-dominated spectral profile and a cluster-dominated profile occurs near 3700~\AA, the canonical cross-over wavelength for LRDs \citep{setton}.  The correspondence between the observed LRDs and the globular cluster in formation model is particularly remarkable considering the SMS model was \textit{not} tuned to the observed LRD properties, rather the model was developed before LRDs were discovered as a distinct class of objects.

The globular cluster in formation model also matches multi-wavelength LRD observations. The SMS-only model has a 4050-to-3600~\AA\ $F_\nu$ flux density ratio of 5.1, matching the strongest observed LRDs \citep{naidu25, degraff_cliff}. An SMS without corresponding UV emission would resemble a \lq{}\lq{}black-hole star\rq{}\rq{}. Using the inferred $A_{1500}$, the observed \muv\ of $-$19.4~mag \citep{torralba}, and energy conservation, we estimate $L_{\rm IR} \sim 4\times10^{11}$~L$_\odot$, fully consistent with the observed upper-limit for A2744-45924 of $L_{\rm IR} < 1.6\times10^{12}$~L$_\odot$  \citep{setton}. The SMS model has an approximately Rayleigh-Jeans tail, suggesting it should be faint in the MIR \citep{setton, degraff_cliff}. Finally, SMSs are too cool to produce significant X-ray emission or ionizing photons \citep{martins}. The nebular emission lines are likely powered by the young stellar population instead of the SMS. Overall, the multi-wavelength LRD properties are consistent with a model of a globular cluster forming with an active SMS. 

\subsection{Potential Supermassive Star Properties}\label{sms}
\begin{figure}
    \centering
    \includegraphics[width=\linewidth]{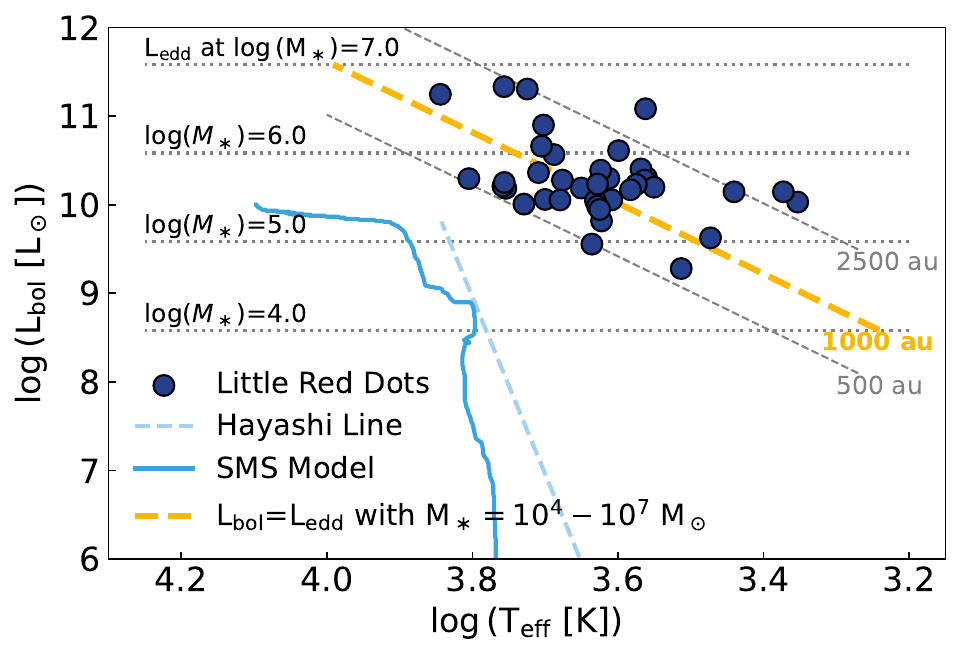}
    \caption{Comparison of the effective temperature ($T_{\rm eff}$) and the bolometric luminosity ($L_{\rm{bol}}$) as measured from modified blackbody fits to a sample of Little Red Dots \citep[blue point;][]{degraaf}. The solid light-blue line illustrates a model for the evolution of an accreting supermassive star in hydrostatic equilibrium \citep{nandal}. The SMS model is significantly cooler and less luminous than the LRDs, possibly because it does not account for a dense wind (\autoref{sms}). The blue-dashed line shows the Hayashi Line, or the maximum luminosity of a fully convective star in hydrostatic equilibrium. In dashed gray, we show lines of constant radius (500 and 2,500~au), while the dotted gray horizontal lines show the Eddington Luminosity of an object with $\log($\mstar/M$_\odot) = 4, 5, 6,$ and 7. The gold-dashed line shows the Eddington Luminosity for a $1,000$~au object with mass increasing from $10^4$ to 10$^7$~M$_\odot$. The Little Red Dot population is consistent with a $10^{4-7}$~M$_\odot$ star emitting at, or above, the Eddington Luminosity. }
    \label{fig:hr}
\end{figure}

The globular cluster in formation model in \autoref{fig:spec} does not perfectly reproduce the observed LRDs, particularly near $3500$~\AA\ where the spectrum transitions from cluster to SMS dominated. These discrepancies may indicate (1) the SMS model is too hot, or (2) the SMS winds are insufficiently dense. Cooler blackbodies peaking near $5,000$~\AA\ may match the data better. However, CMFGEN \citep{hillier} does not include molecular opacities (e.g., H$^-$), and no SMS atmosphere model cooler than $7,000$~K currently exists. An optically thick SMS wind would provide its own opacity that may explain the cooler observed LRD temperatures. Thus, the observed discrepancies may constrain the stellar astrophysics of SMSs rather than rule out the proposed scenario. 

\autoref{fig:hr} compares the $T_{\rm eff}$ versus bolometric luminosity ($L_{\rm bol}$) from modified blackbody fits to the rest-frame optical \citep{degraaf}. LRDs occupy a region of cooler temperatures and higher luminosities than the current SMS models \citep[solid blue curve; ][]{nandal}. LRDs also lie above the Hayashi Line \citep[][]{kippenhahn}, which sets the maximum luminosity of a fully-convective star in hydrostatic equilibrium. For H$^-$ opacity with [Fe/H]$ = -1.5$, the Hayashi Line has a slope $\frac{d\log T_{\rm eff}}{d\log L_{\rm bol}} \sim 0.05$. The SMS model runs parallel to the Hayashi Line because it assumes hydrostatic equilibrium \citep{nandal}. The high radiation pressure instead favors an outwardly expanding outer envelope, with photospheres set by the wind physics rather than interior structure.

Instead, LRDs are consistent with emission from an extended envelope produced by an optically thick continuum-driven wind \citep{dotan}. The Eddington Luminosity is
\begin{equation}
    L_{\rm edd} = \frac{4 \pi G m_{\rm p} c}{\sigma_{\rm T}} M = 3.3 \times 10^4 ~L_\odot \left(\frac{M_\ast}{M_\odot}\right),  \label{eq:ledd}
\end{equation}
where $G$ is the gravitational constant, $m_{\rm p}$ is the proton mass, $c$ is the speed of light, and $\sigma_{\rm T}$ is the Thomson scattering cross section. Electron scattering is likely to be the dominant opacity interior to the photosphere, and H$^-$ opacity can dominate at the cooler temperatures exterior to the photosphere \citep{gray, kippenhahn}. LRDs with $L_{\rm bol} \gtrsim L_{\rm edd}$ will launch electron-scattering-driven winds. The SMS $L_{\rm bol}-T_{\rm eff}$ relation is estimated by equating $L_{\rm edd}$ to the Stefan-Boltzmann law as
\begin{equation}
    L_{\rm edd} = 4 \pi R_{\rm ph}^2 \sigma T_{\rm eff}^4 ,
\end{equation}
where $\sigma$ is the Stefan-Boltzmann constant. The gold-dashed line in \autoref{fig:hr} shows this relation for a fixed photospheric radius $R_{\rm ph} = 1,000$~au and $4\le\log($\mstar/M$_\odot)\le7$. The LRDs scatter about this relation, consistent with emission from an extended wind-driven photosphere. $R_{\rm ph}$ is order-of-magnitude consistent with the sonic point of an isothermal electron-scattering-driven wind from a 10$^{5.5}$~M$_\odot$ star at $L/L_{\rm edd} \sim 0.99$ \citep[][]{lamers_and_cassinelli, dotan}. These inflated photospheres plausibly explain the cooler $T_{\rm eff}$ and higher $L_{\rm bol}$ of LRDs relative to SMS models in hydrostatic equilibrium (\autoref{fig:hr}). 

If such SMS winds are responsible for the abundance patterns observed in local globular clusters, they must escape the stellar gravitational potential. For $\log($\mstar$) = 5.5$  and $R_{\rm ph} = 1,000$~au, the escape velocity is $>750$~km~s$^{-1}$. Additional broadening mechanism  \citep[e.g. electron scattering; ][]{rusakov, chang} can further broaden the wind profiles to produce the broad Balmer lines observed in LRDs. The rest-frame optical emission from LRDs is consistent with arising from the photosphere of an optically thick radiation-driven wind, plausibly alleviating tensions between SMS models and LRD observations. Improved SMS stellar wind and atmosphere models, that account for molecular opacities, are required to test this scenario.

\section{LRD Number Counts}
\label{LRDs} 
Having shown that the V-shaped spectral morphology of LRDs is plausibly reproduced by a globular cluster in formation with an active SMS, we now turn to the observed LRD luminosity function and number counts. In the next two sub-sections, we first use the $z\sim5-7$ UV luminosity function to predict the shape of the $z\sim0$ mass function, adopting the deliberately simple assumption of a constant cluster mass-loss rate from $z\sim7$ to $z\sim0$ (\autoref{shape}). We then estimate the total number of LRDs formed over the entire observed redshift range (\autoref{counts}). 
\begin{figure}[t!]
\includegraphics[width=\linewidth]{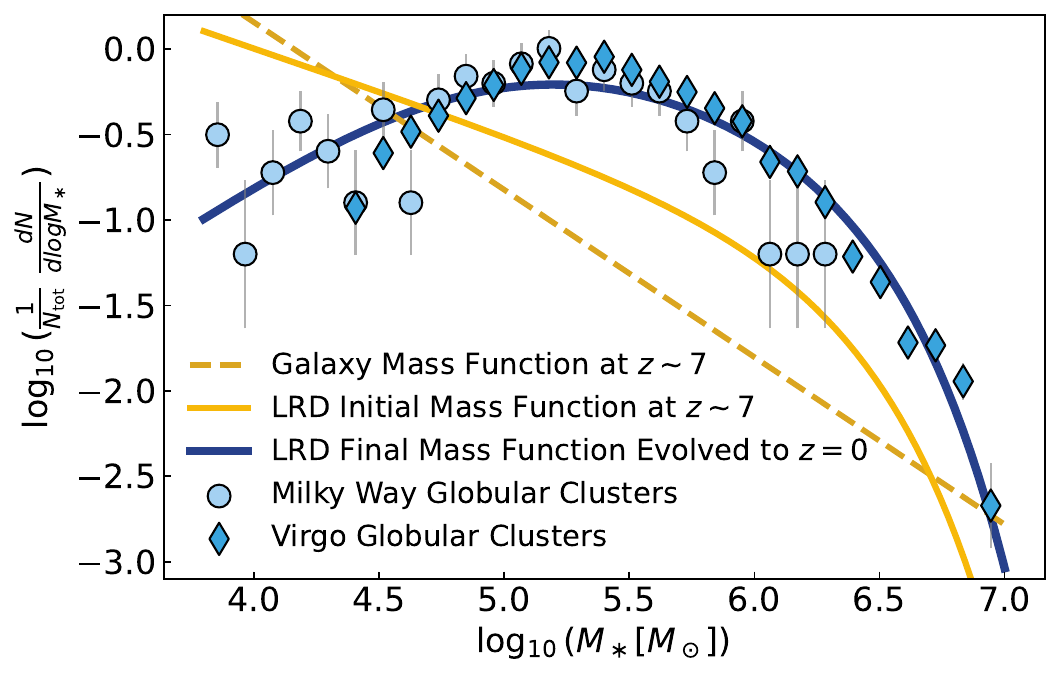}
\caption{The estimated evolution of the LRD mass function from $z\sim7$ (solid gold line) to $z\sim0$ (blue curve). We use the observed LRD UV luminosity function \citep{kokorev24} and an extreme UV mass-to-light ratio that is consistent with a young globular cluster with an active supermassive star. We compare the predicted $z\sim0$ mass function to globular clusters in the Milky Way \citep[light-blue circles; ][]{harris} and in Virgo \citep[dark-blue diamonds;][]{jordan}. The extrapolation of the $z\sim7$ galaxy stellar mass function \citep[dashed gold line;][]{navarro} to these \mstar\ has a  distinct shape from the LRDs. All mass functions have been normalized by the total number to emphasize their shapes. LRDs can plausibly match the observed local globular cluster mass function using standard assumptions about their mass-loss (see \autoref{shape}). 
\label{fig:lf}}
\end{figure}

\subsection{The present-day shape of the LRD mass function}\label{shape}
The LRD UV luminosity function (UVLF) is constructed using photometric samples selected for compactness and V-shaped spectral profiles \citep{kocevski, kokorev24}. A Schechter function \citep{Schechter} matches the LRD UVLF, with minimal redshift evolution between $z\sim 5$ and $z\sim7$ \citep{kokorev24}. We adopt the LRD UVLF from \citet{kokorev24} with a power-law slope $\alpha = -1.5\pm0.4$, a characteristic \muv\ of $M^\ast = -20.7\pm0.7$~mag, and a normalization of $\phi^\ast = (5\pm 2)\times10^{-6}$~Mpc$^{-3}$.

Consistent with the interpretation of LRDs as globular clusters (see \autoref{formation}), many LRDs are associated with extended companion galaxies \citep{rinaldi25, zhang25}.  We conservatively assume 20\% of the total UV luminosity arises from the compact object, and that the SMS contributes 20\% of this compact emission \citep{wendy}. Thus, the cluster light is $\sim16$\% of the total UV luminosity.

We convert the UV luminosity into \mstar\ assuming that LRDs are globular clusters in formation. We adopt an age of $\sim1$~Myr, motivated by the short lifetimes of SMSs  \citep{gieles18}. Forming an SMS requires clusters populating the extreme upper-end of the IMF ($\gtrsim100$~M$_\odot$), implying a UV continuum dominated by very massive stars ($\gtrsim150$~M$_\odot$) with low mass-to-light ratios \citep[e.g.,][]{sabhahit}. We adopt stellar population synthesis models appropriate for this situation  \citep[][]{schaerer25, Hawcroft} extending to $\sim400$~M$_\odot$ with 1~Myr age, nebular continuum included, metallicity of 20\%~$Z/Z_\odot$, and a high-mass IMF slope of $\alpha = -2$, yielding $\frac{M_\ast}{L_{\rm UV}} = \frac{10^6 M_\odot}{2\times10^6~\text{L}_\odot \,\text{\AA}^{-1}}$. Lower-metallicity models (Z/Z$_\odot\sim3$\%) that include rotation may further reduce this mass-to-light ratio \citep{Hawcroft}. While extreme, this mass-to-light ratio naturally arises from the proposed scenario of a young globular cluster in formation.

The resulting LRD mass function at $z\sim 7$ is shown as the solid gold curve in \autoref{fig:lf}. We normalize by the total number of objects to emphasize its shape and enable a comparison to local objects. The inferred LRD initial mass function has a classic Schechter function shape: an exponential cut-off above $\log$(\mstar/$M_\odot$)~$\sim 6$ and a steep power-law slope extending to lower masses.

Most LRDs have blue continua with $\langle\beta_{\rm UV}\rangle \approx - 1.8$ and weak $\beta_{\rm UV}$-\muv\ trends \citep[][]{Matthee, degraaf}. Dust attenuation shifts the inferred mass functions by $0.1-0.5$~dex, but does not significantly change the shape unless there is a  $\beta_{\rm UV}$-\muv\ relation. The dust correction is therefore degenerate with uncertainties of the observed LRD luminosity function, and we do not apply a correction.

Once young clusters form, how do they evolve? After the LRD phase, we assume the SMS exhausts its core hydrogen and collapses to a black hole \citep{begelman, woods}. The cluster then ceases being identified as an LRD because the SMS is no longer producing rest-frame optical continuum. The remaining massive stars continue to produce blue continua until they explode as supernovae after 3-10~Myr. This stellar feedback clears the remaining dust and gas from the cluster to shut off star formation \citep{turner}. The cluster then passively evolves until $z\sim0$.

Over the ensuing $\sim13$~Gyr, the quenched cluster loses \mstar\ through stellar evolution and dynamical interactions such as tidal stripping and evaporation \citep{ostriker, spitzer, fall, lamers}. We adopt the simplest approach of a total constant mass-loss of 10$^5$~M$_\odot$ ($\approx10$~M$_\odot$ Myr$^{-1}$), consistent with recent analyses of Milky Way stellar streams and N-body simulations \citep{gieles23, chen25, chen25a}. We tested more complex mass-loss prescriptions, such as a \mstar-dependent mass-loss \citep[with $\gamma = 2/3$; ][]{lamers}, and find a largely unchanged shape to the final mass function.

The $z\sim7$ LRD mass function evolved to $z\sim0$ with constant mass-loss is shown as the blue curve in \autoref{fig:lf}. The exponential cut-off for $>10^6$~M$_\odot$ remains at $z\sim0$ because the mass-loss does not disrupt the most massive clusters, whereas the lowest mass clusters are dissipated. The surviving number of clusters depends on the initial low-mass power-law slope because it sets the abundance of clusters initially near the destruction mass. Consequently, the steep initial mass function develops a turnover near $\log($\mstar/M$_\odot) \sim 5.3$.

For comparison, \autoref{fig:lf} also shows the $z\sim7$ galaxy stellar mass function \citep{weaver, navarro}. At the \mstar\ relevant for globular cluster formation, the extrapolated galaxy stellar mass function is a single power law. Without an exponential cutoff at $>$10$^{6}$~M$_\odot$, it does not naturally evolve into the same shape as the LRD scenario.

\subsection{Total number density of LRDs at $z\sim0$}\label{counts}
The previous section focused on the shape of the mass function;  we now estimate the total number density of LRDs formed. We estimate the instantaneous comoving number density of observed LRDs, $n_{\rm inst}$, by integrating the observed $z\sim7$ UVLF, $\phi^{\rm UV}$,  as 
\begin{equation}
    n_{\rm inst} = \int_{-23}^{-15} \phi^{\rm UV} dM_{\rm UV}   \approx 1\times10^{-4}~\text{Mpc}^{-3}, 
\end{equation}
where the integration limits correspond to $\log($\mstar/M$_\odot) = 4-7$, the mass range included in \autoref{fig:lf}. Lower-mass values are unlikely to survive dynamical interactions down to $z\sim0$.

SMS lifetimes set the LRD observability window and are estimated to be on the order of 1~Myr. The LRD formation rate is then
\begin{equation}
    R = \frac{n_{\rm inst}}{t_{\rm obs}} \approx 0.1~\text{Mpc}^{-3}~\text{Gyr}^{-1} . 
\end{equation}
The total number of LRDs formed between two redshift intervals, $z_{\rm min}$ and $z_{\rm max}$, assuming a constant $R$, is then
\begin{equation}
    n_{\rm tot} = \int_{z_{\rm min}}^{z_{\rm max}} R \frac{dt}{dz} dz  \approx  0.3~\text{Mpc}^{-3}  \label{eq:ntot} ,
\end{equation}
where we used $z_{\rm min} = 2$ and $z_{\rm max} = 10$ to match the observed LRD redshift range \citep{kocevski, kokorev24, taylor}. 

These number densities are order of magnitude estimates. \autoref{eq:ntot} assumes that all LRDs survive to $z\sim0$, but clusters could be dynamically dissipated. For example, if only 30\% of the LRDs survive until $z\sim0$, we estimate an observed number density of $\approx0.1$~Mpc$^{-3}$ at $z\sim0$. With reasonable assumptions, the total number density of LRDs formed across all redshifts is $\approx0.1-0.3$~Mpc$^{-3}$.

\section{Redshift zero globular clusters}
\label{GCs}
Now that we have predicted the LRD mass function at $z\sim 0$, we can compare it to observed globular clusters around the Milky Way \citep{harris}, Andromeda \citep{usher24}, and the Virgo cluster \citep{jordan}. These three systems are extremely well studied and span a large range of \mstar. We adopt a constant M$_\ast$/L$_V = 2$ for Virgo and 1.5 for Milky Way and Andromeda, respectively, consistent with an old stellar population \citep{mclaughlin, jordan07}. We use the reported distances for the Milky Way and Andromeda globular clusters, and a distance modulus of 31.09 for Virgo \citep{jordan}. Number densities are determined in 0.15~dex \mstar\ bins with Poisson uncertainties. 

The Milky Way and Virgo mass functions are shown as blue circles and squares in \autoref{fig:lf} (Andromeda follows a similar shape). The globular cluster mass function peaks near $\log($\mstar/M$_\odot) = 5.3$, declines exponentially to higher masses, and has a power-law decline to lower masses, consistent with previous results \citep{harris91, brodie06, jordan07}. The local globular clusters have a similar mass function shape as predicted  for $z\sim0$ LRDs.  

We also estimate the total number of globular clusters in the local Universe. This depends both on the number density of clusters per halo and the number density of host halos harboring the globular clusters \citep{spitler, harris13, harris15, mbk17}. The Milky Way is roughly an $L^\ast$ galaxy with an abundance of $n_{\rm L^\ast}\sim 0.005$~Mpc$^{-3}$ \citep{Blanton} and has $N_{\rm gc} \approx 150$ globular clusters. This implies
\begin{equation}
    n_{\rm GC} = N_{\rm gc} n_{\rm L^\ast} \approx 0.8~\text{Mpc}^{-3},
\end{equation}
consistent with literature estimates of $0.5-1.5$~Mpc$^{-3}$ \citep{harris13, harris15, mbk17}. This order-of-magnitude globular cluster number density is similar to the LRD number density (\autoref{eq:ntot}). Both the shape of the LRD mass function and the  total number counts are consistent with present-day globular cluster populations. 


\section{Predictions if Little Red Dots are Globular Clusters in Formation}
\label{discuss}
The previous three sections have shown it is plausible that LRDs are globular clusters in formation. Here, we conclude with testable predictions of this hypothesis using known globular cluster properties. These concrete predictions provide an opportunity to stress-test the hypothesis to hopefully gain a better understanding of the true nature of LRDs.

\begin{figure}
    \centering
    \includegraphics[width=\linewidth]{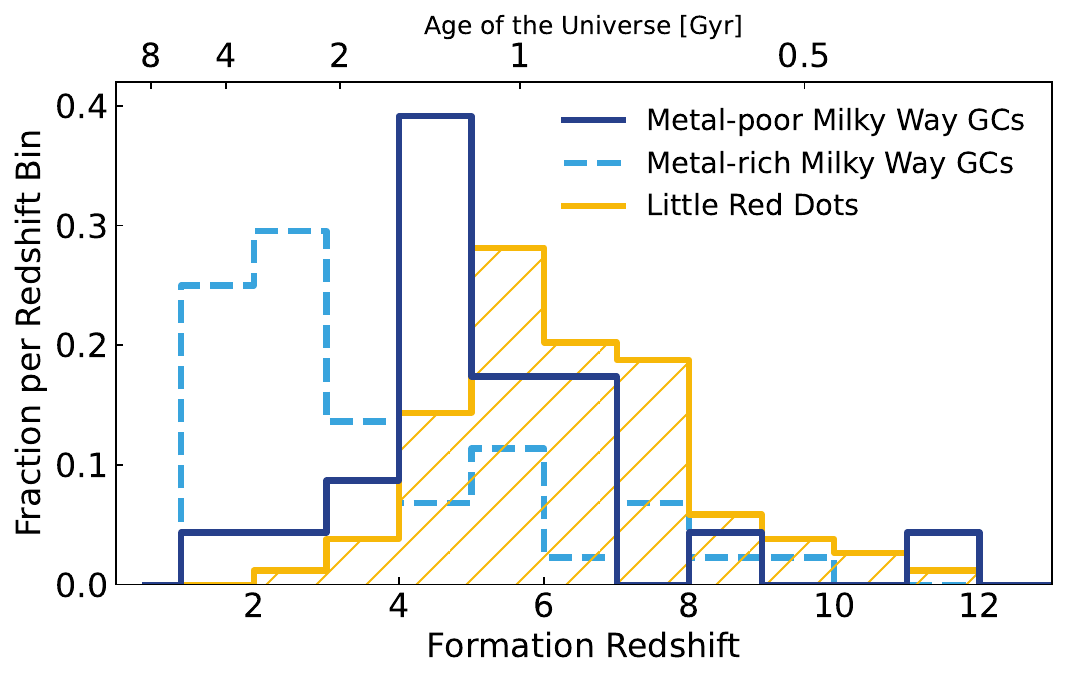}
    \caption{Comparison of the formation redshifts for Little Red Dots \citep[gold hatched; ][]{kocevski} to globular clusters in the Milky Way \citep[][]{forbes}. We conservatively split the Milky Way globular clusters as metal-rich ([Fe/H] $>-1.4$; dashed light-blue) and  metal-poor ([Fe/H] $< -1.4$; dark-blue line) to emphasize the distinct populations. The formation redshift of metal-poor Milky Way globular clusters roughly matches the redshift distribution of LRDs. }
    \label{fig:age}
\end{figure}

\subsection{LRDs may represent metal-poor globular clusters}\label{formation}

Globular clusters are generally quite old, with observations and models indicating a mean formation epoch at $z \gtrsim 2$ and formation rates peaking near $z \sim 4-6$ \citep{cohen, kruijssen15, forbes, bastian, elbadry, kruijssen19, ying, Aguado-Agelet, valenzuela}. If LRDs trace globular cluster formation, they should form primarily at these epochs.  

This formation timescale roughly matches the observed LRD redshift range of $z\sim 4-8$  \citep{kocevski, kokorev24, ma25}. In \autoref{fig:age}, the LRD peak \citep{kocevski} aligns with the old, metal-poor Milky Way globular cluster population \citep[dark blue; ][]{forbes}. In contrast, the metal-rich globular clusters (light-blue dashed line) lack a corresponding LRD population at $z\sim1-2$.  Metal-rich globular clusters are typically found within $\sim 1R_e$ of galactic bulges, and show weaker evidence of multiple stellar populations. This suggests that either a population of $z\sim1-2$ LRDs remains undetected in dense, metal-rich environments, or that SMS formation is suppressed at higher metallicities.

LRDs frequently have nearby neighbors \citep{Matthee, labbe24}, with $\approx40$\% within $\approx 1$~kpc of a neighbor \citep{baggen}. This resembles the bimodal spatial distribution of local globular clusters and naturally explains the spatially extended UV emission as host galaxy contamination \citep{zhang25, rinaldi25, wendy}. By analogy with local globular clusters, LRDs forming within 1~$R_e$ of more massive companions should be more metal-enriched than isolated systems. A currently undetected population of $z\sim1-2$ LRDs forming contemporaneously with galactic bulges could therefore account for the missing metal-rich globular cluster population. 

Alternatively, SMS formation may depend on metallicity.  Metal-rich Milky Way globular clusters that formed at lower redshift tend to show lower stellar surface densities \citep{forbes}. This trend likely reflects not only metallicity, but also ISM pressures and environmental conditions which shape cluster collapse and structure \citep{kruijssen15}. If SMS formation requires high stellar densities to drive runaway stellar collisions \citep{gieles18}, then lower density clusters forming at lower redshift would be less likely to reach sufficient densities to form SMSs. In this scenario, low-redshift globular clusters would not produce SMSs and would lack the characteristic V-shaped spectral profile required for LRD selection.  

The redshift distribution provides an empirical test for the LRD-globular clusters hypothesis. The observed LRD peak aligns with metal-poor globular clusters, while no LRD population comparable to the metal-rich systems are seen at lower redshift. This absence may reflect either an observational bias against detecting them in dense, metal-rich environments or a metallicity ceiling for SMS formation.

\subsection{Some LRDs should have high He abundances and anti-correlations between Na-O, Al-Mg, and N-C }\label{abundances}
Globular clusters have distinct, and oftentimes multiple, abundance patterns due to neutron capture \citep{Gratton}. Many show helium-enhanced populations \citep{gratton10, dupree}, along with anti-correlations between odd and even atomic number elements, including: Na-enhanced and O-poor, Al-enhanced and Mg-poor, and N-enhanced and C-poor populations \citep{bastian, milone}. Although not universal, the trends are most prominent in the most massive \citep{carretta} and oldest \citep{bastian} globular clusters. If LRDs represent globular clusters in formation, similar chemical signatures may be present in a subset of LRDs. 

These abundance patterns are challenging to diagnose in LRDs due to limited spectral resolution and the lack of sufficient SMS models. \ion{He}{1} emission lines are among the strongest lines detected in nearly all LRDs \citep{Matthee, labbe24, kokorev26, torralba, wendy}, with strengths comparable to hydrogen lines. While a detailed He/H abundance has not yet been reported, SMSs nucleosynthesis models predict enhanced helium abundance. 

The rest-frame optical contains strong neutral sodium transitions (Na~D 5890, 5896~\AA\AA), while the rest-UV has resonant aluminum lines (\ion{Al}{2} 1670~\AA\ and \ion{Al}{3} 1855, 1863~\AA\AA). However, few LRD spectra have sufficient resolution to isolate these lines \citep{eugenio}. A2744-45924, shown in \autoref{fig:spec} \citep[][]{labbe24}, has a tentative \ion{Al}{3}~1860~\AA\ detection  and relatively weak \ion{Mg}{2}~2800~\AA\ emission blended with \ion{Fe}{2} lines. The Al, Mg, and Fe lines could provide a direct comparison to local globular clusters, with [Al/Mg]$> 0$ supporting the globular cluster in formation picture \citep{carretta, bastian, milone}. 

Carbon emission is reported in several LRDs \citep{akins25,  tang, akins26}, whereas nitrogen is less common because broad H$\alpha$ contaminates the [\ion{N}{2}]~6549, 6584\AA\AA\ lines. Alternatively, the UV high-ionization lines may directly probe the nitrogen abundance \citep[e.g.,][]{labbe24}. A recent systematic search in the Dawn JWST Archive identified high-ionization nitrogen lines in six LRDs \citep{morel}. These objects lack UV carbon lines, and have weak rest-optical oxygen lines, suggesting a tentative N-C and N-O anti-correlation. However, the current evidence for distinct abundances patterns in LRDs remains weak and relies on sparse samples. A systematic study of the relative abundances is required to robustly test the globular cluster hypothesis. 

\subsection{Should metal-poor globular clusters host intermediate mass black holes?}\label{dyn}
Globular clusters have long been modeled with numerical simulations and evolutionary models to match their surface brightness and velocity dispersion profiles. A central SMS remnant, most plausibly a black hole,  would increase the central velocity dispersion and delay core collapse. \citet{Baumgardt} modeled local globular clusters including central black holes and derived upper-limits of  $\log($M$_{\rm BH}/M_\odot) \leq 4$ for most local globular clusters.  Observationally, whether globular clusters host black holes is actively debated \citep{mezcua, greene20}. $\omega$ Centauri is the most likely candidate \citep{Noyola, haberle}, while most others remain undetected. Local globular clusters put stringent mass constraints on SMS remnants.

What black hole mass is expected from an SMS? If LRDs radiate near $L_{\rm edd}$, the inferred SMS masses could reach $\log($\mstar/M$_\odot) \approx 6.5$, potentially exceeding the dynamical upper-limits in local globular clusters. However, SMS winds may remove $\gtrsim 90$\% of the SMS mass prior to collapse \citep{dotan, Denissenkov}, substantially reducing the resulting black hole mass. Such mass loss also provides a natural resolution to the multiple abundances populations \citep{gieles18}. Improved SMS atmosphere and wind models are needed to determine whether optically thick winds yield remnant masses consistent with those inferred from local globular clusters.

\section{Conclusion}
We have tested the hypothesis of Little Red Dots as globular clusters in formation with an active supermassive star. This scenario is motivated by the legacy of local globular cluster observations and provides a physically coherent framework for interpreting LRD observations. While the scenario remains plausible, no definitive conclusion can be currently made and we outline several observations to stress-test this hypothesis. If confirmed, Little Red Dots offer a direct view of globular cluster formation and open a new window onto extreme stellar astrophysics characterized by intense radiation fields. Their high luminosities suggest that similar systems may be observable at even earlier epochs, potentially characterizing the first generations of stars.

\begin{acknowledgments}
JC thanks Hollis Akins, Volker Bromm,  Rui Chaves-Marques, Steve Finkelstein, Karl Gebhardt, Keith Hawkins, Harley Katz, Stellar Offner, Daniel Schaerer, Grace Telford, and Jorick Vink for conversations that improved the paper.
AdG acknowledges support from a Clay Fellowship awarded by the Smithsonian Astrophysical Observatory.
MBK acknowledges support from NSF grants AST-2108962 and AST-2408247; NASA grant 80NSSC22K0827; HST-GO-16686, HST-AR-17028, JWST-GO-03788, and JWST-AR-06278 from the Space Telescope Science Institute, which is operated by AURA, Inc., under NASA contract NAS5-26555; and from the Samuel T. and Fern Yanagisawa Regents Professorship in Astronomy at UT Austin.
AACS acknowledges support by the Deutsche Forschungsgemeinschaft (DFG, German Research Foundation) in the form of an Emmy Noether Research Group -- Project-ID 445674056 (SA4064/1-1, PI Sander). AACS further acknowledges support from the Deutsches Zentrum f\"ur Luft und Raumfahrt (DLR) grant grants 50 OR 2509 (PI: A.A.C.\ Sander) and 50 OR 2306 (PI: V.\ Ramachandran/A.A.C.\ Sander) as well as from the Federal Ministry of Research, Technology, and Space (BMFTR) and the Baden-Württemberg Ministry of Science as part of the Excellence Strategy of the German Federal and State Governments. This project was co-funded by the European Union (Project 101183150 - OCEANS).
\end{acknowledgments}

\begin{contribution}


\end{contribution}

%
\facilities{JWST(NIRSpec)}



\appendix


\bibliography{lrd}{}
\bibliographystyle{aasjournalv7}



\end{document}